\documentstyle[prl,twocolumn,aps]{revtex}

\begin{document}

\draft

\title{Microwave Resonance and Carrier-Carrier Interaction in
 Two-Dimensional Hole Systems at High Magnetic Field}
\author{C.-C. Li, J. Yoon,
L. W. Engel$^{\dag}$, D. Shahar$^{*}$, D. C. Tsui, and M. Shayegan}
\address{Department of Electrical Engineering, Princeton University,
Princeton, NJ 08544}

\address{$^{\dag}$National High Magnetic Field Laboratory, Florida
State University, Tallahassee, FL 32306} \address{$^{*}$Department of
Condensed Matter Physics, Weizmann Institute, Rehovot 76100, Israel}

\maketitle

\begin{abstract}
Microwave frequency conductivity, Re($\sigma_{xx}$), of high quality
two-dimensional hole systems (2DHS) in a large perpendicular magnetic
field ($B$) is measured with the carrier density ($n_s$) of the 2DHS
controlled by a backgate bias.
The high $B$ insulating phase of  the 2DHS
exhibits a microwave resonance  that
remains well-defined, but shifts to higher peak frequency (\mbox{$f_{pk}$})
as $n_s$ is reduced.  Over a wide
range of $n_s$, \mbox{$f_{pk}$} $\propto n_s^{-1/2}$ is
observed for the two   samples we measured.
The data clearly indicate
that both carrier-carrier interactions and disorder are indispensable
in determining the dynamics of the insulator.  The $n_s$ dependence of
 $f_{pk}$ is consistent with   a weakly pinned Wigner crystal
in which  domain size increases with $n_s$, due to larger
carrier-carrier interaction.

\end{abstract}

\narrowtext

In high magnetic field, $B$, sufficiently high quality two dimensional
systems of electrons or holes (2DES or 2DHS) in semiconductors exhibit
the fractional quantum Hall effect (FQHE)\cite{fqhe}, a spectacular
manifestation of carrier-carrier interaction.  In the high $B$ limit,
samples that show the FQHE become insulators, and it is natural to ask
what is the role of the carrier-carrier interaction in this insulating
phase.  The fields and carrier densities at which the high $B$
insulating phase appears in FQHE-exhibiting samples are in rough
agreement with predictions \cite{wcpredict} of the appearance of a
Wigner crystal (WC), stabilized by the freezing out of the kinetic
energy in the magnetic field. In a WC ground state, carriers form a
lattice, to minimize the energy of their mutual repulsion, so
carrier-carrier interaction is of central importance.  For a WC to be
an insulator, disorder is required, which pins the WC, and also causes
the WC to have finite correlation length, or domain size ($L$).  On
the other hand, if disorder is too strong, the carrier-impurity
interaction can localize carriers resulting in an insulator without
Wigner crystalline order.  Hence the central questions about the high
$B$ insulating phases pertain to a competition between carrier-carrier
and carrier-impurity interactions.

The picture of a pinned WC motivated many different types of
experiments
\cite{vladdepin,lidepin,williams,lidiel,liinduct,cyclotron,photolum,SAW,M124,prl97,mellorep2dsjapan}
on the high $B$ insulator, including conduction
\cite{vladdepin,lidepin,williams}, noise generation
\cite{vladdepin,lidepin}, giant dielectric constant \cite{lidiel},
ac-dc interference \cite{liinduct}, cyclotron resonance
\cite{cyclotron}, and photoluminesence measurements \cite{photolum}.
Microwave measurements
\cite{williams,SAW,M124,prl97,mellorep2dsjapan}, such as those
presented here, reveal resonant absorption in the high $B$ insulators
of samples capable of exhibiting the FQHE. Such resonances are
typically interpreted as ``pinning modes'', in which WC domains
oscillate within the pinning potential of the disorder.

Theories due to Fukuyama and Lee (FL)\cite{fl}, and later
investigators \cite{nlmandml} describe the pinning mode as WC domains
oscillating in parabolic restoring potentials. The potentials are
proportional to $\omega_0^2$, where $\omega_0$ is called the pinning
frequency.  In the magnetic field, such oscillators have two modes.
The higher mode frequency, $\omega_+$, is shifted above the cyclotron
frequency $\omega_c=eB/m^*$, where $m^*$ is the carrier effective
mass.  If $\omega_0\ll\omega_c$, the lower mode frequency, $\omega_-$,
is $\omega_0^2/\omega_c$, and is identified with
$2\pi$\mbox{$f_{pk}$}, where \mbox{$f_{pk}$} is the measured resonance
peak frequency.  It is important to note that the $m^*$ cancels in the
expression for \mbox{$f_{pk}$}.  In units of \mbox{Re$(\sigma_{xx})$}
integrated over $f$, FL obtain oscillator strengths of $S =n_s\pi
f_{pk} e/B$ for the lower mode.  Recent measurements\cite{prl97} on
high quality 2DHS show resonances that depend on $B$ in a way the FL
theory cannot readily explain and that are too sharp to be explained
easily as independently oscillating, disorder-induced domains. As yet,
a real understanding of the static configuration of the insulator, as
well as its dynamics, is lacking.

 In this paper we directly address the centrally important competition
 between carrier-carrier and carrier-impurity interaction, by varying
 carrier density $n_s$ in a sample, without warming it up. This
 effectively varies the carrier-carrier interactions while leaving the
 disorder potential unchanged.  We present data on the microwave
 resonance of the high $B$ insulating phase of high quality 2DHS
 samples, measured with varying hole density ($n_s$) by means of a dc
 bias voltage applied to a backgate. In a constant $B$ of 13 T, the
 resonance peak remains well-defined, with decreasing peak height, as
 $n_s$ is reduced.  The evolution of the resonance peak with
 decreasing $n_s$ shows a true shift of center frequency,
 \mbox{$f_{pk}$}, not just a lopsided $f$-dependent reduction of
 intensity.  This change in oscillator center frequency must be due to
 carrier-carrier interaction, and we can reject any picture of the
 resonance with purely independent carriers.  Interpretation of the
 decreasing \mbox{$f_{pk}$} vs $n_s$ in terms of a pinned WC requires
 a ``weak pinning'' model, in which domain size $L$ is determined by a
 balance between the carrier-carrier interaction and carrier-impurity
 interaction energies\cite{fl}, and in which $L$ increases with
 increasing $n_s$.  Over a wide range in $n_s$, for the samples we
 examined,  \mbox{$f_{pk}$} $\propto$ $n_s^{-1/2}$, in agreement with
 theory\cite{fl,nlmandml} for fixed $B$.

The microwave technique for measurement of Re($\sigma_{xx}$) was just
as in refs. \cite{M124} and \cite{prl97}.  \mbox{Re$(\sigma_{xx})$}
was obtained from the attenuation of a microwave transmission line,
patterned onto the top of the sample, as shown in the inset to Fig.~1,
and coupled capacitively to the 2DHS.  No spatial harmonics were
observed, so the experiment is sensitive to $\sigma_{xx}$ with
wavevector $q\lesssim 2\pi/W$, where $W= 30\ \mu$m is the width of gap
between center and side conductors of the line. Hence we take the
\mbox{Re$(\sigma_{xx})$} data presented as characteristic of the 2DHS
in the long wavelength limit.  The real part of the diagonal
conductivity is calculated from transmitted power $P$ as in
refs. \cite{M124,prl97}, Re$(\sigma_{xx})$=$W\vert \ln
P\vert$/$2Z_0d$, where $d$=28 mm is the total length of the
transmission line.  $Z_0=50 \Omega$ and is the transmission line
characteristic impedance when $n_s=0$, and $P$ is normalized to unity
for that condition.  Errors of this formula are estimated as in
refs. \cite{M124,prl97}, as 15 percent, with the apparatus typically
20 times more sensitive to \mbox{Re(\mbox{$\sigma_{xx}$})} than it is
to Im(\mbox{$\sigma_{xx}$})

We present data on two p-type samples (1 and 2), from wafers also used
  by Santos {\em et al.}\ \cite{RIP} in an observation of re-entrance
  of insulating behavior around the 1/3 FQHE.  Sample 1 was from the
  same wafer used in an earlier microwave study\cite{prl97} of the
  high $B$ insulator.  Samples 1 and 2 are both
  GaAs/Al$_{x}$Ga$_{1-x}$As heterostructures grown on (311)A GaAs
  substrate, with Si modulation doping in two layers.  The $n_s$ of
  each sample was reduced by means of a bias voltage applied to a
  backgate, and measured at various positive backgate bias voltages
  from the quantum Hall effect.  Increasing $n_s$ with negative bias
  was impractical due to parallel conduction.  At zero bias, sample 1
  had $n_s\sim 5.5\times 10^{10}$ cm$^{-2}$, and mobility $3.5 \times
  10^{5}$ cm$^2$/V-s, while sample 2 had $n_s\sim 4.2\times 10^{10}$
  cm$^{-2}$, and mobility $5.0 \times 10^{5}$ cm$^2$/V-s.  Details of
  the growth of the samples are similar; parameters of the nearest
  doping layer to the 2DHS may be relevant to the data we will
  present: sample 1 ( sample 2) has 1150 \AA\ (1200\AA) of
  Al$_{0.35}$Ga$_{0.65}$As (Al$_{0.3}$Ga$_{0.7}$As) between the 2DHS
  and the doping layer, of thickness 44 \AA (46 \AA) and 2D density
  9$\times 10^{11}$cm$^{-2}$ (8$\times 10^{11}$cm$^{-2}$).

Fig.~1 shows the 0.2 GHz Re($\sigma_{xx}$) vs $B$ for sample 1 at zero
backgate bias.  For this scan and all other measurements presented in
this paper the temperature of the samples was $T\sim$ 25 mK.  A series
of well-developed FQHE conductivity minima can be seen, and
Re($\sigma_{xx}$) vanishes again for $B$ $>$ 10 T, for the system is
well inside the insulating phase.  These features are consistent with
those of dc magnetoresistance data.

Fig.~2 shows Re($\sigma_{xx}$) vs $f$ traces of sample 1 for several
$n_s$'s in $B$ = 13 T.  On reducing $n_s$, $f_{pk}$ increases by more
than a factor of 3 before Re($\sigma_{xx}$) diminishes to such an
extent that the resonant maximum cannot be clearly
identified. Decreasing $n_s$ also broadens the peak and reduces its
maximal \mbox{Re$(\sigma_{xx})$}.  The error in reading $n_s$ from the
dc transport data is estimated to be $\sim 3$\% for high $n_s$ and
$\sim$ 10\% for low $n_s$.  The small rounded feature at the shoulder
of the $n_s=5.42\times 10^{10}$cm$^{-2}$ peak is instrumental
artefact.

Fig.~3 summarizes parameters of the resonance for the two samples,
measured as $n_s$ varies, in fixed $B$ of 13 T.  Fig.~3a shows
\mbox{$f_{pk}$} vs $n_s$ from two different cooldowns of each of
\mbox{sample 1} and \mbox{sample 2}.  For a given sample {$f_{pk}$}
varies only 10 percent between cooldowns.  Variation between the two
samples is much larger, and so is apparently intrinsic to the wafers.
In both samples the $B$ dependence of \mbox{$f_{pk}$}, even at reduced
$n_s$, is much like that observed in ref. \cite{prl97}:
\mbox{$f_{pk}$} vs $B$ increases, but flattens out at large enough
$B$.  By 13 T, \mbox{$f_{pk}$} vs $B$ is in this flat regime, so that
the 13 T \mbox{$f_{pk}$}, plotted in Fig.~3a, can be regarded as an
approximate high $B$ limiting value.

The \mbox{$f_{pk}$} vs $n_s$ behavior is simpler for sample 2, though
the measured range of $n_s$ is less than it was for sample 1.  Sample
2 exhibits \mbox{$f_{pk}$} uniformly decreasing with $n_s$, with fits
of the data to the power law \mbox{$f_{pk}$}$\propto n_s^{-\gamma}$
that are consistent, given the errors and range of measurement, with
$\gamma=1/2$.  Least squares fits, shown on the graph as solid lines,
result in $\gamma=0.46$ and 0.51 for the two different cooldowns of
sample 2 (shown as $\bigcirc$ and \raisebox{-2pt}{ {\large $\Box$}}).
The \mbox{$f_{pk}$} vs $n_s$ traces for Sample 1 show three different
regions of $n_s$. In the highest $n_s$ region, ($n_s > 5.0\times
10^{10} $ cm$^{-2}$), \mbox{$f_{pk}$} changes little with $n_s$. In
the intermediate $n_s$ region, ($3.2\times 10^{10}\le n_s \le 5.0
\times 10^{10}$ cm$^{-2}$), the data can be fit well to
\mbox{$f_{pk}$}$\propto n_s^{-3/2}$, and least squares fits to
\mbox{$f_{pk}$}$\propto n_s^{-\gamma}$ result in the dotted lines
shown in the figure, for which $\gamma=1.66$ and 1.24 for the two
cooldowns (shown as \raisebox{-2pt}{ {\huge $\bullet$} } and
\rule{1.5ex}{1.5ex} ).  In the low $n_s$ region, $1.34\times 10^{10}
<n_s<3.2\times 10^{10}$ cm$^{-2}$, the data on sample 1 are, like all
\mbox{$f_{pk}$} vs $n_s$ for sample 2, consistent with
\mbox{$f_{pk}$}$\propto n_s^{-1/2}$.  The least squares fit lines
shown in Fig.~3a give $\gamma=0.58$ for both cooldowns of sample 1 in
the low $n_s$ region.

Fig.~3b shows $S/$\mbox{$f_{pk}$} vs $n_s$ for the two samples. $S$ is
the numerical integral of \mbox{Re$(\sigma_{xx})$} vs $f$ taken over
the experimental $f$ range of 0.2 to 6.0 GHz.  $S$ is a good measure
of the resonance oscillator strength when the tails of the resonance
do not extend beyond this $f$ range.  $S/$\mbox{$f_{pk}$} is plotted
in Fig.~3b for comparison with the FL oscillator
model\cite{fl,nlmandml} prediction that $S/$\mbox{$f_{pk}$}$=n_s\pi
e/B$.  While the $S/$\mbox{$f_{pk}$} data do appear directly
proportional to $n_s$, the slope of the fit line drawn in the figure,
is 0.47 of the FL predicted value, $\pi e/B$.  The linearity of
$S/$\mbox{$f_{pk}$} vs $n_s$ suggests the sum rule can be generalized
to a realistic model of the resonance; disagreement with the FL
prediction for the slope is expected considering the inconsistency of
that model with the observed $B$ dependence of the resonance as
reported in ref. \cite{prl97}. The measured value of the slope is
consistent with the resonance oscillator strengths reported
earlier\cite{M124,prl97}, which are all roughly half of the FL values.

The evolution of the peak as $n_s$ is reduced leads to the definite
conclusion that the resonance cannot be modeled as noninteracting
individual carriers bound to defects in the semiconductor host.
Fig.~2 shows that as $n_s$ is reduced, \mbox{Re$(\sigma_{xx})$}
increases with decreasing $n_s$ in a region of the high $f$ wing of
the peak, though the integrated conductivity $S$ decreases.  For
example, \mbox{Re$(\sigma_{xx})$} at 2.55 GHz increases from 1.2 to 13
$\mu$S, on reducing $n_s$ from 5.42 to 3.13 $\times 10^{10}$
cm$^{-2}$.  This means reducing $n_s$ does not only remove oscillators
responsible for the resonance, but changes the parameters of these
oscillators.  This can only happen when the carrier-carrier
interaction plays a role in determining \mbox{$f_{pk}$}.

 The data also allow us to rule out ``strong pinning'' by dilute
impurities. In such a model, the pinning potential is strong enough to
essentially immobilize carriers bound to impurities, so that rigidly
confined domains between impurities oscillate, and the domain size $L$
is independent of the stiffness of the crystal. As $n_s$ is increased,
the mass density of the 2D system increases, as does the stiffness of
the WC.  We maintain that an increase of mass density does not slow
down oscillation through inertia for modes with frequency
$2\pi$\mbox{$f_{pk}$}$\ll\omega_c$. Low frequency
($\omega(q)\ll\omega_c$) modes\cite{bm} of the disorder-free classical
WC are independent of $m^*$ and hence of mass density; such modes are
taken to give the pinning mode frequency when $q\sim 1/L$. Likewise,
in high $B$ a harmonic oscillator made up of part of the 2D system has
pinning mode frequency $\omega_-$ as set down in ref. \cite{fl},
independent of the oscillator mass density.  The case of strong
pinning is contrary to what is observed since in that extreme,
increasing $n_s$ would increase the WC stiffness and leave $L$
unchanged, and so would increase \mbox{$f_{pk}$}.

We interpret the increase of \mbox{$f_{pk}$} with decreasing $n_s$ as
a general consequence of weak WC pinning. By definition, in the weak
pinning regime, the forces between carriers in the WC are larger than
the forces exerted on carriers by impurities, and the Lee-Rice
correlation length, or domain size, $L$, is larger than the WC lattice
constant, $a$. The lattice is crystalline over length $L$, so the
random impurity potential is averaged over area $L^2$, resulting in
the reduction of the effective pinning as $L/a$ increases.  Increasing
$n_s$ increases WC stiffness and $L/a$, causing carriers to tend to
stay closer to their crystalline positions and to ``fall'' less into
the impurity potential, so that the restoring force and
\mbox{$f_{pk}$} are reduced as $L$ is increased.

The weak pinning theory in Refs. \cite{fl} and \cite{nlmandml} is a
 well-developed example of how weak pinning causes \mbox{$f_{pk}$} to
 decrease with $n_s$ and even agrees with the \mbox{$f_{pk}$}$\sim
 n_s^{-1/2}$ behavior exhibited by much of the data.  Following
 ref. \cite{nlmandml}, $\omega_0$ is the same as the frequency of a
 transverse mode of the clean, $B=0$ WC at wavevector $q=1/L$:
 $\omega_0^2=\mu/{n_s}m^*L^2$, where $\mu$ is the shear modulus of the
 classical WC, $\mu=\alpha n_s^2e^2a/(4\pi\epsilon_0\epsilon)\sim
 n_s^{3/2}$, and $\alpha\approx 0.02$.  The $n_s^{-1/2}$ behavior of
 $\omega_0^2$ (and hence \mbox{$f_{pk}$}) is seen on evaluating $L$ by
 minimizing the total energy, $L\sim \mu a^2/n_i^{1/2} V_0 \sim
 n_s^{1/2}$, where $n_i$ is the two- dimensional impurity density, and
 $V_0$ is the impurity potential strength.

While we conclude that a weak pinning mechanism is needed to explain
the observed decrease of \mbox{$f_{pk}$} with $n_s$, we know of no
theory as yet that can explain all observations on the resonance.The
theory of refs. \cite{fl,nlmandml} gives $L/a $ comparable to unity
for observed \mbox{$f_{pk}$}'s, contrary to the expectation for weak
pinning conditions, and also does not explain \cite{prl97} the $B$
dependence or sharpness at larger $n_s$ of the resonances.  By taking
into account incommensurate pinning enhancement of the pinning
frequency, the work of Ferconi and Vignale\cite{fv} gives much larger
$L/a$ for the observed \mbox{$f_{pk}$}. Chitra {\em et
al.}\cite{chitra} compute conductivity spectra for a weakly pinned WC;
the results depend on a number of length scales that characterize the
pinned WC. The theory of Fertig\cite{fertig} predicts the sharpness of
the resonances as a consequence of the long-range of the Coulomb
interaction, and can address the observed $B$
dependence\cite{M124,prl97} of the resonance.  Fertig's theory can
also be cast in a weak pinning picture to give \mbox{$f_{pk}$}$\sim
n_s^{-3/2}$, as observed for sample 1 for $5.0\times10^{10} > n_s>
3.2\times10^{10}$ cm$^{-2}$.

The observed differences between the two samples must be explained by
their different disorder.  In these samples, the sources of disorder
are remote acceptors, impurities (mainly C acceptors), and
heterojunction interface characteristics.  By most indications, the
disorder of sample 1 is more significant than that of sample 2. In
sample 1, smaller spacer and larger doping would increase the
influence of remote acceptors, and its mobility and FQHE definition
are less than in sample 2.  At the same $n_s$ and $B$, larger disorder
would give a larger restoring force in weak pinning, so a larger
\mbox{$f_{pk}$} is expected, consistent with the \mbox{$f_{pk}$} of
sample 1 always exceeding that in sample 2.  One possible explanation
for the two regimes of decreasing \mbox{$f_{pk}$} vs $n_s$ in sample 1
would be that more than one type of disorder is playing a role in
determining \mbox{$f_{pk}$}.

In summary, we studied the $n_s$ dependence of the microwave
 resonance in the high $B$ insulating phase of
high quality 2DHS samples, and observed an upward shift in \mbox{$f_{pk}$}
as $n_s$ was reduced.  Neither single-particle
localization nor an ideal strongly pinned WC
could produce such a dependence.  When
interpreted as the pinning mode of a 2D WC, the data point to weakly
pinned WC domains, whose size increases with $n_s$.

We thank R. Chitra, H. A. Fertig, M. Hilke, D. A. Huse, and
P. Platzman for helpful discussions. This work was supported by NSF,
and L.W.E.  acknowledges support from the state of Florida.

 \newpage
\centerline{FIGURES}

FIG.~1.\ \ Real part of 0.2 GHz diagonal conductivity vs magnetic
field of sample 1 at $T$ $\sim$ 25 mK and zero backgate bias.  The inset
shows the transmission line on sample surface, with black indicating
the evaporated Al film.

 \ \\

FIG.~2.\ \ Real part of diagonal conductivity vs frequency of sample 1
 at various carrier densities, $n_s$, in a constant magnetic field of 13 T,
 $T$ $\sim$ 25 mK.

\ \\

FIG.~3.\ \ a.\ \ Resonance frequency vs hole density for both
\mbox{sample 1} (\raisebox{-2pt}{ {\huge $\bullet$} } and
\rule{1.5ex}{1.5ex} ) and \mbox{sample 2} (shown as $\bigcirc$ and
\raisebox{-2pt}{ {\large $\Box$}}) in a constant magnetic field of 13
T and $T$ $\sim$ 25 mK.  Different symbols for the same sample denote
different cooldowns.  Both solid lines and dotted lines are power law
least squares fits to the data.  \ b. \ Ratio of integrated
Re$(\sigma_{xx})$ over resonance frequency \mbox{$f_{pk}$}, vs hole
density.  The solid line is a linear least squares fit to the data.

\end{document}